\documentstyle[epsf,bibnorm,float,amsbsy]{lamuphys}
\makeatletter
\let\chapter\hid@chapter
\makeatother
\begin{document}
\pagenumbering{arabic}

\title{Neutron-Proton Interaction in Doubly Odd Deformed Nuclei}

\author{Nunzio Itaco, Aldo Covello, and Angela Gargano}

\institute{ Dipartimento di Scienze Fisiche, Universit\`a di Napoli
Federico II, and Istituto Nazionale di Fisica Nucleare, Complesso Universitario
di Monte S. Angelo, Via Cintia, I-80126 Napoli, Italy}

\maketitle

\begin{abstract}
In this paper, we present some results of a particle-rotor model study
of $^{176}$Lu. In particular, we consider the two lowest $K^{\pi} = 1^+$ 
bands, which 
present a rather large odd-even staggering. This effect, which may
be traced to direct Coriolis coupling with Newby-shifted $K^{\pi} =0^+$ bands,
is of great interest 
since it gives information on the neutron-proton interaction.
We use both zero-range and finite-range interactions with
particular attention focused on the role of the tensor force. 
Comparison of the calculated
results with experimental data evidences the importance of the tensor-force 
effects in the description of the odd-even staggering in $K \ne 0$ bands.
\end{abstract}

\section{Introduction}
It is well known that the two most important effects associated with 
the residual neutron-proton interaction in doubly odd deformed nuclei are the 
Gallagher-Moszkowski (GM) splitting \cite{Gallagher58} and the odd-even or
Newby (N) shift \cite{Newby62}.
Further information on the
neutron-proton interaction may be obtained by studying the odd-even staggering
\cite{Jain89} in $K \ne 0$ bands.
In fact, the main mechanism responsible for this effect is the mixing, 
through the Coriolis interaction, of $K \ne 0$ bands with one or more
N-shifted $K = 0$ bands. Bands which exhibit odd-even 
staggering represent therefore an indirect source of knowledge of 
the effective neutron-proton interaction.

In a previous paper \cite{Covello97}, we have performed a complete
Coriolis band-mixing calculation for the doubly odd deformed nucleus $^{176}$Lu 
within the framework of the particle-rotor
model.
Our aim was to assess the role of the effective neutron-proton
interaction, with particular attention focused on the tensor force.
Here, we report on our results concerning the two lowest $K^{\pi} = 1^+$ 
bands in $^{176}$Lu,
which exhibit a rather large odd-even staggering.

The paper is organized as follows. In Sec. 2 we give a brief description
of the model and some details of our calculations. In Sec. 3 we compare 
our results with the experimental data. Some concluding remarks are given in Sec. 4.

\section{Outline of the model and calculations}
We assume that the unpaired neutron and proton are strongly coupled
to an axially symmetric core and interact through an effective 
interaction.
The total Hamiltonian is written as
\begin{equation}
H=H_0 + H_{{\mathrm RPC}} + H_{{\mathrm ppc}} + V_{np}.
\end{equation}
The term $H_0$ includes the rotational energy of the whole system, 
the deformed, axially 
symmetric field for the neutron and proton, and the intrinsic contribution
from the rotational degrees of freedom.
It reads
\begin{equation}
H_0 = \frac{\hbar^2}{2 {\cal J}} ( \boldsymbol {I}^2 - I_3^2 ) + H_n + H_p +
\frac{\hbar^2}{2 {\cal J}} [ ( \boldsymbol {j}_n^2 - j_{n3}^2 ) + 
( \boldsymbol {j}_p^2 - j_{p3}^2 ) ].
\end{equation}
The two terms $H_{{\mathrm RPC}}$ and $H_{{\mathrm ppc}}$ stand for 
the Coriolis coupling
and the coupling of particle degrees of freedom through the rotational motion,
respectively. Their explicit expressions are
\begin{equation}
H_{{\mathrm RPC}} = - \frac{\hbar^2}{2 {\cal J}} (I^+J^- + I^-J^+),
\end{equation}
\begin{equation}
H_{{\mathrm ppc}} = \frac{\hbar^2}{2 {\cal J}} (j_n^+j_p^- + j_n^-j_p^+).
\end{equation}
The effective neutron-proton interaction is given the general form
\begin{equation}
V_{np}=V(r) [ u_0 + u_1 \boldsymbol {\sigma}_p \cdot \boldsymbol {\sigma}_n + 
u_2 P_M +u_3 P_M \boldsymbol {\sigma}_p \cdot \boldsymbol {\sigma}_n 
+ V_T S_{12} + V_{TM} P_M S_{12} ],
\end{equation}
with standard notation \cite{Boisson76}.
In our calculations we have used a finite-range force with a radial
dependence $V(r)$ of the Gaussian form 
\begin{equation}
V(r) = {\mathrm exp} (-r^2/r_0^2),
\end{equation}
as well as a zero-range force.
In the latter case, $V_{np}$ takes the simple form
\begin{equation}
V_{np}^{\delta} = \delta (r) [ v_0 + v_1 \boldsymbol {\sigma}_p \cdot 
\boldsymbol {\sigma}_n ].
\end{equation}
As basis states we use the eigenvectors of $H_0$ properly symmetrized 
and normalized,
\[
| \nu _n \Omega_n \nu_p \Omega_p I M K \rangle =  
\left ( \frac{2I + 1}{16 \pi^2} 
\right )^{\frac{1}{2}} [ D^I_{MK} | \nu _n \Omega_n \rangle | \nu_p \Omega_p
\rangle 
\]
\begin{equation}
~~~~~~~~~~~~~~~~~~~~~~~~~~ + (-)^{I+K} D^I_{M-K}  
|\nu_n \overline{\Omega}_n \rangle | \nu_p \overline{\Omega}_p \rangle ], 
\end{equation}
where the state $| \nu \overline{\Omega} \rangle$ is the time-reversal
partner of $| \nu \Omega \rangle$. 

We have used the standard Nilsson potential \cite{Gustafson67} to generate the 
single-particle Hamiltonians $H_n$ and $H_p$. The parameters $\mu$ and $\kappa$
have been obtained from the mass-dependent formulas of Ref. \cite{Nilsson69}.
As regards the deformation parameter $\beta_2$ we have used the value 0.26,
according to Ref. \cite{Elmore76}.

The single-particle energies for the odd proton and the odd
neutron and the rotational parameter $\hbar^2 /2 {\cal J}$ 
have been derived from the experimental spectra \cite{Macchiavelli93} 
of the two neighboring
odd-mass nuclei $^{175}$Yb and $^{175}$Lu.
The explicit values of the parameters involved in the calculation
may be found in Ref. \cite{Covello97}.

For the neutron-proton interaction, as already mentioned above,
 we have used both a finite-range force with
a Gaussian radial shape and a zero-range interaction.
As regards the former,
we have performed two different calculations, with and without 
the tensor terms.
For the parameters of this interaction we have adopted the values determined
by Boisson {\it et al.} \cite{Boisson76} in their analysis of
the GM splittings and N shifts in the rare-earth region.
As regards the $\delta$ force, no value of the strength of the spin-spin term
$v_1$ gives a satisfactory description of the N shifts. 
We have used the value $v_1 = -0.20 $ MeV, which leads to the lowest 
possible disagreement between theory and experiment.
More details about the choice of the parameters of the potentials as well as
their explicit values are given in Ref. \cite{Covello97}.

\section{Results and comparison with experiment}
In the low-energy spectrum of $^{176}$Lu two $K^{\pi} = 1^+$ bands
have been observed, which both exhibit a rather large 
odd-even staggering \cite{Jain89}. 
\begin{figure}[h]
\centerline{
\epsfbox{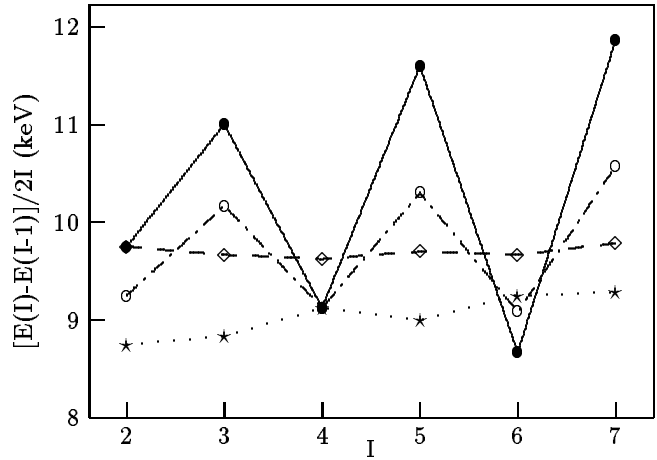}
}
\caption[]{Experimental and calculated odd-even staggering of the
$K ^{\pi} =1^+ p \frac{9}{2}[514] n \frac{7}{2}[514]$ band in $^{176}$Lu.
Solid circles correspond to experimental data. The theoretical results
are represented by open circles (Gaussian central plus tensor force),
diamonds (Gaussian central force), and stars ($\delta$ force).}
\end{figure}

They are the $K ^{\pi} =1^+ p 
\frac{9}{2}[514] n \frac{7}{2}[514]$ band starting at 197 keV, and
the $K ^{\pi} =1^+ p \frac{9}{2}[624] n \frac{7}{2}[404]$ band, whose bandhead
lies at 339 keV.

The experimental ratio \cite{Klay91} $[ E(I) - E(I-1)] /2I$ for the 
lowest $K^{\pi} = 1^+$ band is plotted
vs $I$ and compared with the calculated ones in Fig. 1.
We see that the experimental behavior is satisfactorily reproduced
by the calculation including the tensor force.
When using the pure central Gaussian force the staggering is almost 
nonexistent and in the case of the $\delta$ force not only is its magnitude
very small, but it has also the opposite phase.
To better understand our results let us discuss the structure of 
the states of this $K^{\pi} = 1^+$ band.
In all of our three calculations we find that the wave functions 
of these states contain significant components (5 -- 10 \%)
with $K^{\pi} = 0^+ p \frac{7}{2} [523]
n \frac{7}{2} [514]$. The mixing between the two bands is obviously
due to the Coriolis interaction, whose effects are 
enhanced by the presence of the $\frac{9}{2} [514]$
and $\frac{7}{2} [523]$ single-proton orbitals arising from the 
$h_{11/2}$ shell-model state.  
The fact that only the calculation including the tensor terms gives the right
staggering implies that only this force is able to reproduce a sizable N shift
for the $K^{\pi} = 0^+ p \frac{7}{2} [523] n \frac{7}{2} [514]$ band.
Unfortunately, this band has not been definitely recognized, so 
that a direct comparison is not possible at present.

Let us now come to the $K^{\pi} =1^+ p \frac{7}{2} [404] n\frac{9}{2} [624] $
band.
In Fig. 2 we plot the experimental ratio \cite{Klay91}
$ [E(I) - E(I-1)]/2I$ 
vs $I$ and compare it with the calculated ones.
\begin{figure}[H]
\centerline{
\epsfbox{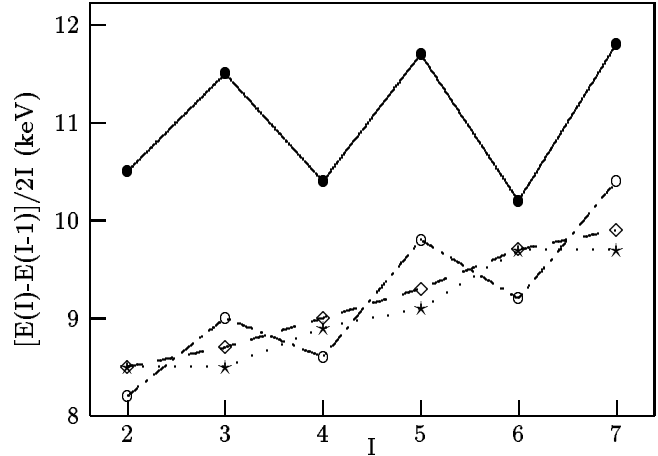}  
}
\caption[ ]{Same as Fig. 1, but for the $K^{\pi} =1^+ p \frac{7}{2} [404]
n \frac{9}{2} [624] $ band.}
\end{figure}

We see that also in this case only the calculation including the tensor force is able 
to reproduce the correct experimental odd-even staggering, the main discrepancy
being a downshift of the calculated value of about 2 keV.
The analysis of the wave functions of the states of this $K^{\pi}= 1^+$
band shows that it is considerably mixed with the $K^{\pi} = 0^+
p \frac{7}{2} [404] n \frac{7}{2} [633]$ band.
In fact, in all of our three calculations the wave functions contain 
significant components (5 -- 15 \%) with $K^{\pi} = 0^+$.
While this band has not been observed in $^{176}$Lu, it has been 
recognized in $^{174}$Lu  
with a N shift of $-44$ 
keV \cite{Browne91}.
For the latter band we find a N shift of $-26$, $1$, and $6$ keV, 
using a Gaussian force with tensor terms, a pure central Gaussian force, and 
a $\delta$ force, respectively.
This provides evidence that the odd-even staggering in the $K^{\pi} = 1^+$
band is to be traced to tensor force effects in the description 
of the N-shifted 
$K^{\pi} = 0^+ p \frac{7}{2} [404] n \frac{7}{2} [633]$ band. 

\section{Concluding remarks}
In this paper, we have presented the results of a study of the two
lowest $K^{\pi} = 1^+$ bands in $^{176}$Lu within the framework of the 
particle-rotor model. Our aim was to assess the role of the effective
neutron-proton interaction in the description of $K \ne 0$ bands
in doubly odd deformed nuclei.
In Ref. [4] we showed that the space-exchange
and spin-spin space-exchange forces as well as the tensor force
are very relevant in 
the description of the N shifts in $K = 0$ bands.
The results of the present work evidences the role of these forces 
in the description of some $K \ne 0$ bands. In particular, 
we show that, owing to the Coriolis coupling, the odd-even staggering
observed in the two lowest $K^{\pi} = 1^+$ bands in $^{176}$Lu
appears to be a manifestation of the tensor force effects.
In conclusion, we feel that it is well worth extending the present
study to other $K \ne 0$ bands which exhibit a rather large
odd-even staggering.
Work in this direction is in progress.

%
%
%


\end{document}